*Field-induced antiferromagnetic correlations in a nanopatterned van der Waals ferromagnet: a potential artificial spin ice*

*Avia Noah*[*1,2,3], *Nofar Fridman*[1,2], *Yishay Zur*[1,2], *Maya Markman*[1], *Yotam Katz King*[1,2], *Maya Klang*[1], *Ricardo Rama-Eiroa*[),5], *Harshvardhan Solanki*[5], *Michael L. Reichenberg Ashby*[1,6], *Tamar Levin*[1], *Edwin Herrera*[7], *Martin E. Huber*[8], *Snir Gazit*[1], *Elton J. G. Santos**[4,5,9], *Hermann Suderow*[7], *Hadar Steinberg*[1,2], *Oded Millo*[1,2], *and Yonathan Anahory**[1,2]

[1]The Racah Institute of Physics, The Hebrew University, Jerusalem, 9190401, Israel
[2]Center for Nanoscience and Nanotechnology, Hebrew University of Jerusalem, Jerusalem, 91904, Israel
[3]Faculty of Engineering, Ruppin Academic Center, Emek-Hefer, 40250 Monash, Israel
[4]Donostia International Physics Center (DIPC), 20018 Donostia-San Sebastián, Basque Country, Spain
[5]Institute for Condensed Matter Physics and Complex Systems, School of Physics and Astronomy, University of Edinburgh, Edinburgh, EH93FD, United Kingdom
[6]Imperial College London, Blackett Laboratory, London, SW7 2AZ, United Kingdom
[7]Laboratorio de Bajas Temperaturas, Unidad Asociada UAM/CSIC, Departamento de Física de la Materia Condensada, Instituto Nicolás Cabrera and Condensed Matter Physics Center (IFIMAC), Universidad Autónoma de Madrid, E-28049 Madrid, Spain
[8]Departments of Physics and Electrical Engineering, University of Colorado Denver, Denver, CO 80217, USA
[9]Higgs Centre for Theoretical Physics, University of Edinburgh, Edinburgh EH93FD, United Kingdom

Email: avia.noah@mail.huji.ac.il, esantos@ed.ac.uk, yonathan.anahory@mail.huji.ac.il

**Abstract**

Nano-patterned magnetic materials have opened new venues on the investigation of strongly correlated phenomena including artificial spin-ice systems, geometric frustration, magnetic monopoles, for technologically important applications such as reconfigurable ferromagnetism. With the advent of atomically thin two-dimensional (2D) van der Waals (vdW) magnets a pertinent question is whether such compounds could make their way into this realm where interactions can be tailored so that unconventional states of matter could be assessed. Here we show that square islands of $CrGeTe_3$ vdW ferromagnets distributed in a grid manifest antiferromagnetic correlations, essential to enable frustration resulting in an artificial spin-ice. By using a combination of SQUID-on-tip microscopy, focused ion beam lithography, and atomistic spin dynamic simulations, we show that pristine, isolated CGT flakes as small as $150 \times 150 \times 60$ nm$^3$ have tunable dipole-dipole interactions, which can be precisely controlled by their lateral spacing. There is a crossover between non-interacting islands and significant inter-island anticorrelation depending how they are spatially distributed allowing the creation of complex magnetic patterns not observable at the isolated flakes. Our findings suggest that the cross-talk between the nano-patterned magnets can be explored in the generation of even more complex spin configurations where exotic interactions may be manipulated in an unprecedent way.

## Introduction

Magnetic order is the result of an interplay between magnetic exchange, magnetic anisotropy, Zeeman coupling to an external field, and the dipolar interaction. Although the last is typically the weakest term, the dipolar interaction can become significantly large under certain conditions. For instance, the exchange interaction vanishes when magnetic particles are separated. In addition, the anisotropy energy can be reduced by applying an external magnetic field, which suppresses the anisotropy barrier near the coercive field. In this scenario, the dipolar interaction becomes substantial. Such conditions are theoretically met in single-molecule magnets during the magnetization reversal process, where dipolar interaction is expected to cause magnetic self-ordering[1–3]. The absence of experimental evidence may be partially due to the small spatial scales and the incompatibility of most surface magnetic microscopy techniques with the surface quality of crystallized single-molecule magnets. To circumvent this problem, extensive research has been conducted on larger magnetic particles lithographically patterned from thin films[4–6]. That platform was used to study a broad variety of magnetic lattices such as 1D chains[7], 2D frustrated arrays[8–10], and 3D lattices[11,12]. However, until now this platform was limited to rather simple magnetic thin films such as cobalt or permalloy.

Recent advancements in two-dimensional (2D) materials have unveiled the potential of magnetically ordered van der Waals (vdW) compounds[13–16], with particular interest in the effect of spatial confinement on the magnetic order[17,18]. There have also been recent reports that confinement causes a transition from soft to hard ferromagnetism in $Fe_3GeTe_2$[19], $CrSiTe_3$[20], and $CrGeTe_3$ (CGT)[21]. In addition, ferromagnetism has been observed in atomically thin layers[22] where magnetic anisotropy is indeed reduced substantially. These results are surprising given that magnetism is expected to be weaker in confined structures where, as pointed out by Mermin and Wagner, thermal fluctuations might be expected to suppress long-range ordering in the absence of anisotropy[23]. However, such assumption is only valid in the thermodynamic limit where samples beyond the size of the known universe[24] would hold such dependence of the anisotropy to generate magnetism in 2D. Specifically, it has been experimentally demonstrated that thin CGT films ($d < 10$ nm) exhibit a net magnetization at zero applied field[21,25]. In contrast, the interior of thicker flakes ($d > 10$ nm) has zero net magnetization, with hard ferromagnetism appearing only at the sample edges[21],26. These findings raise the question of the existence of edge effects in the zero-dimension limit, as well as the option that finite size magnets may be manipulated in storage technology applications. It is currently unknown whether nano-flakes of 2D magnets disposed in patterned recording media, *i.e.*, magnetic islands distributed in a grid, could provide any cooperative phenomena useful for manipulating spins in an information bit.

Here, we employ $Ga^+$ FIB amorphization of CGT flakes to fabricate an array of square-shaped magnetic islands with an effective inter-island antiferromagnetic (AFM) interaction which is the essence of artificial spin-ice arrays. Scanning SQUID-on-tip (SOT) microscopy[26,27] is then employed to investigate the magnetic properties of the arrays as a function of the applied field. Our results indicate that the magnetic islands remain fully magnetized at zero applied magnetic field down to the smallest achievable size of $\sim 150 \times 150 \times 60$ nm³, which is surprising considering that a pristine isolated CGT flake of 60 nm thickness does not hold a net magnetization in these conditions[21]. The use of an external field to suppress the anisotropy barrier and reach the coercive field of the magnetic array revealed that the dipolar field causes significant AFM correlations to emerge. Atomistic spin dynamics reproduced closely the effect also indicating that zero-field approaches where the islands are distributed much closer displayed similar anticorrelation features strongly mediated by the stray fields across the magnetic grid. Our results pave the way to study magnetism in nanopatterned arrays of 2D magnets opening a pathway to explore a large number of complex configurations, interactions so far not accessible in vdW materials such artificial spin-ice systems and their vast implications.

## Results

CGT flakes with thicknesses ranging from 30 to 110 nm were exfoliated on top of a SiO$_2$-coated Si wafer. Fig. 1b presents a scanning electron microscope (SEM) image of a CGT flake. Using a 30 keV Ga$^+$ FIB, we amorphized vertical and horizontal lines to fabricate an array of square-shaped magnetic islands (Fig. 1a). Fig. 1e shows a cross-sectional scanning transmission electron microscopy (STEM) image confirming that certain portions of the CGT flake were amorphized (a-CGT, Fig. 1e). As demonstrated in previous work[28], amorphous CGT is nonmagnetic, thereby impeding exchange coupling between the islands. Each magnetic island in the presented array has dimensions of $w \times w \times d = 230 \times 230 \times 30$ nm$^3$ and a separation of $s = 70$ nm. We resolve the local out-of-plane component of the magnetic field, $B_z(x, y)$, of single-islands at 4.2 K (Fig. 1a) by using a scanning SOT microscope (see methods). Figs. 1c,d present SOT images of $B_z(x, y)$ of a magnetic array acquired while applying an external magnetic field of $\mu_0 H_z = 85$ mT, where a black color-coded area indicates islands with a magnetic moment pointing down, while a white color-coded area indicates islands with moments pointing up. The images, which have dimensions of $20 \times 20$ μm$^2$ (Fig. 1c) and $10 \times 10$ μm$^2$ (Fig. 1d) include over 5,000 and 1,000 islands, respectively. The stable magnetic signal demonstrates the magnetic island stability at the nanoscale.

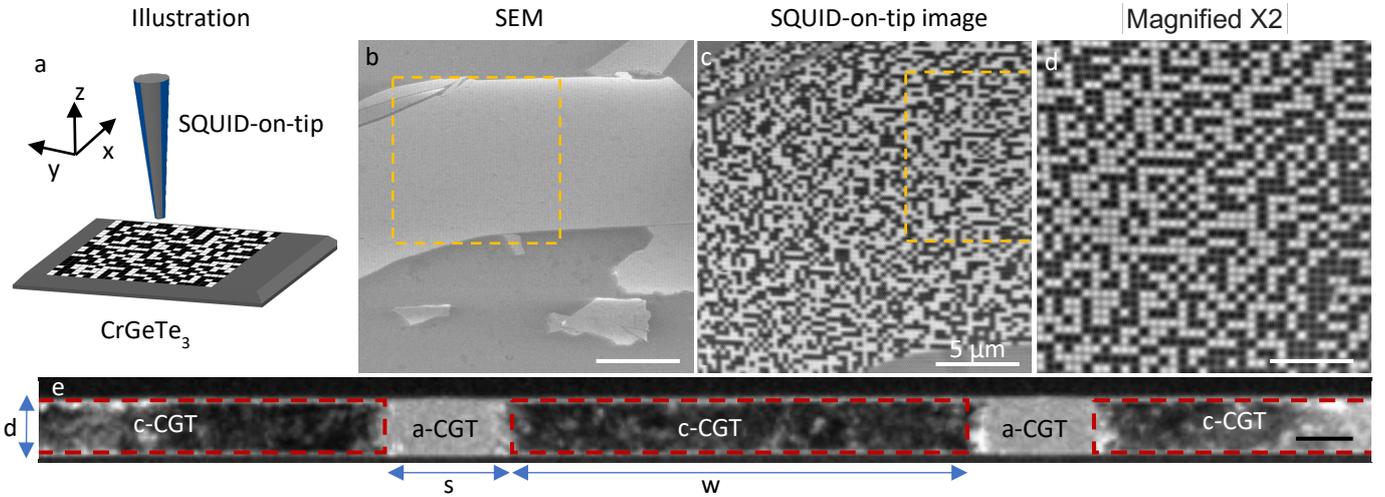

**Figure 1 Stable CrGeTe$_3$ magnetic nanoisland array.** (**a**) Schematic illustration of the experimental setup depicting the SQUID-on-tip (SOT) and the sample. (**b**) Scanning electron microscope image of CrGeTe$_3$ (CGT) patterned using a Ga$^+$ focused ion beam into grid with island size of $230 \times 230 \times 30$ nm$^3$. The area marked with an orange square is shown in **c**. The scale bar is 10 μm. (**c**, **d**) $B_z(x,y)$ SOT images acquired at $\mu_0 H_z = 85$ mT. The area marked with an orange square in **c** is shown in **d**. The SOT images are $20 \times 20$ μm$^2$ **c** and $10 \times 10$ μm$^2$ **d** with a scale bar corresponding to 5 μm **c**, and 2.5 μm **d**. All images were acquired with 35 nm pixel size and acquisition time of 20 min/image. The black/white color scale represents magnetic moments pointing down/up. (**e**) Cross-section scanning transmission electron microscopy of the CrGeTe$_3$ (CGT) etched array. The gray area is amorphous (a-CGT), and the dark area is crystalline (c-CGT). Scale bar is 30 nm.

The response of the patterned CGT array to an applied out-of-plane magnetic field, $H_z$, is shown in Figs. 2a-f, which present a set of $8 \times 8$ μm$^2$ SOT images comprising 529 islands that have been converted into binary images for clarity. The images were acquired at fields ranging between $0 \leq \mu_0 H_z \leq H_s = 150$ mT, following a negative field excursion at $\mu_0 H_z = -200$ mT. The array field evolution is shown in Supplementary Movies 1-2. The SOT image shown in Fig. 2a demonstrates that at $\mu_0 H_z = 0$, each island remains magnetized in the direction of the previously applied field (negative). As $H_z$ increases, the number of islands with magnetization parallel to the field grows at the expense of the number of islands pointing antiparallel to it, until the saturation field, $H_s$, is reached and all the islands point in the positive direction. To quantify the magnetization ($M(H_z)$) of the array, we measured the number of islands oriented in a specific direction. Fig. 2g (red curve) depicts the magnetization curve $M(H_z)$ resulting from the image analysis divided by the magnetic moment of the entire array, $M_{Sat} = n\, m_i$, where $n$ is the number of islands and $m_i$ the magnetic moment of a single island (here $n = 529$ and $m_i \approx 5.7 \times 10^6$ μ$_B$).

The patterned flake possesses the typical hysteresis curve of a hard ferromagnet ($M(H_z = 0) = M_{Sat}$). In contrast, a pristine flake of the same thickness ($d = 30$ nm), exhibits a zero net magnetization at $H_z = 0$, with a bowtie-shaped magnetization curve as measured in previous work[21] (Fig. 2g, blue curve). In that work, hard ferromagnetism was observed at the sample edge. Here, a comparison of the hysteretic loops reveals that confinement has the effect of hardening the ferromagnetism in nanoislands. This comparison highlights the similarity between the magnetic properties of an edge and a nanoisland, which is mainly constituted of edges. The mechanism explaining this unusual edge property remains unknown and will require further investigation. In what follows, we focus on the dipolar interactions between the islands.

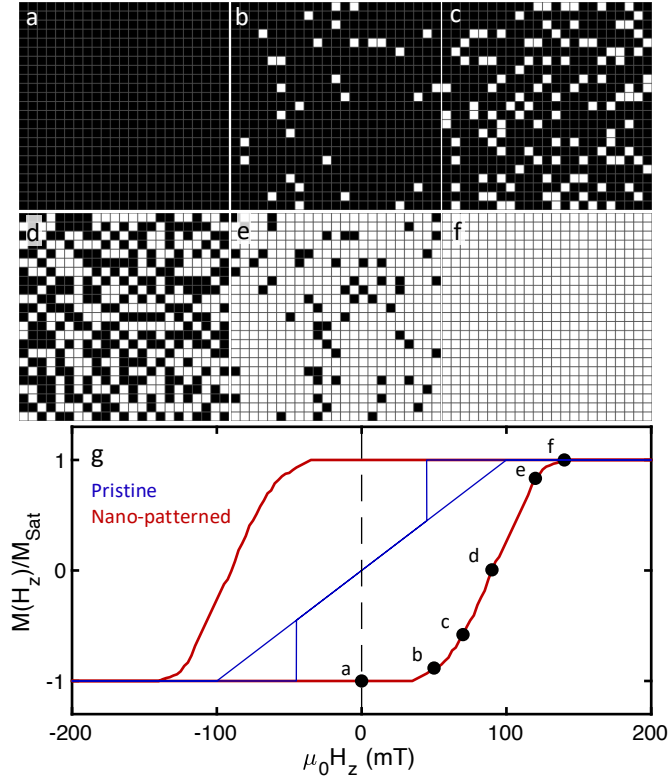

**Figure 2. Magnetic field response of the CrGeTe₃ array magnetization.** (a-f) Sequence of binary matrices computed from the SQUID-on-tip (SOT) $B_z(x,y)$ images at distinct values of an applied out-of-plane field $\mu_0 H_z$. The SOT image sizes are $8 \times 8$ μm². The black/white color scale represents the magnetic moments pointing antiparallel/parallel (down/up) to the applied field. The images were acquired at $\mu_0 H_z = 0$ **a**, 50 **b**, 70 **c**, 90 **d**, 120 **e**, and 150 **f** mT. (**g**) Magnetization curves $M(H_z)$ of nano-patterned (red) and pristine (blue) 30 nm thick CrGeTe₃ flake, in relation to the saturation magnetization, $M_{Sat}$, extracted from the SOT images.

We consider a point dipole in the island generating a dipolar field at the point $\vec{r}$ as illustrated in Fig. 3c. The stray field is given by the following expression $\vec{B}_{dip}(\vec{r}) = \frac{\mu_0}{4\pi}\left[\frac{3\vec{r}(\vec{m}\cdot\vec{r})}{r^5} - \frac{\vec{m}}{r^3}\right]$, where $\vec{r} = (0,0,0)$ is the position vector of the dipole coordinate. In the case of a moment pointing out of the plane and neighbors situated in the plane, we obtain $\vec{r} \perp \vec{m}$, causing the first term of $\vec{B}_{dip}$ to vanish and thus $\vec{B}_{dip} = -\frac{\mu_0 \vec{m}}{r^3}$. The presence of a stray field in the $-\vec{m}$ direction, which couples to the neighboring moment, results in an effective antiferromagnetic interaction between the islands.

Assuming each island is a macrospin $m_i$, we can write the energy of an island as $E_i = K\cos^2\theta - \sum_j E_{dip,i,j} - m_i\mu_0 H_z$, where $K$ is the anisotropy constant and $E_{dip,i,j} = \int \vec{B}_{dip,j}(\vec{r}) \cdot d\vec{m}_i(\vec{r})$ is the energy generated by the dipolar interaction. Here, $\vec{B}_{dip,j}(\vec{r})$ is the integral of the field emanating from all magnetic moments in island $j$ on the infinitesimal element $d\vec{m}_i(\vec{r})$. We compare the expected magnitude of each energy term. The anisotropy constant $K = m_i H_c/2 \gtrsim 10$ eV[29] is estimated considering the island as a macrospin ($m_i = 330$ eV/T) that can either point up or down as our data suggests. At $H_z = 0$, all the arrays remain magnetized in the same direction, suggesting that the magnetic anisotropy term dominates the dipolar interaction (Fig. 3a). This conclusion is reinforced by finite element calculation finding the nearest neighbors $E_{dip} \leq 1$ eV in our geometry. The anisotropy can be reduced significantly by applying an external field near $H_c$ (Fig. 3b). Assuming an arbitrary precision on the applied magnetic field, the energy barrier can be made arbitrarily small, ignoring randomizing effects, such as thermal activation, and assuming that all the islands are identical. Thus, assuming that the dipolar interaction is larger than the randomizing effects at $H_c$, we expect nearest neighbors to point in opposite directions. Further investigations could combine these effective antiferromagnetic interactions, mediated by the dipolar field with other lattices that induce geometric frustration, to study artificial spin ice in magnetic 2D materials. Here, we study out-of-plane magnetization in a square lattice to demonstrate the antiferromagnetic interactions.

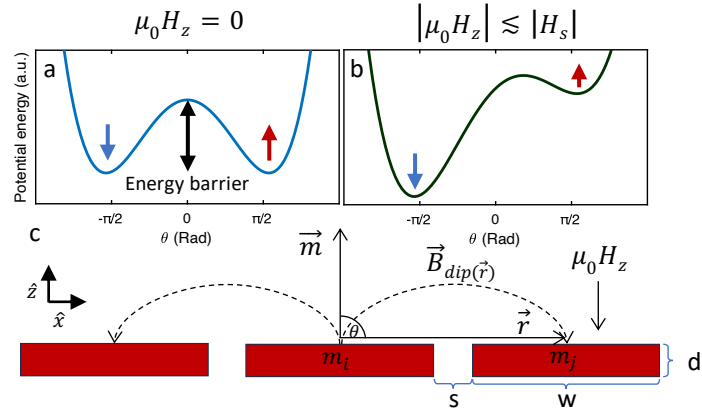

**Figure 3. Energy scales and the dipolar energy $E_{dip}$.** (a) Schematic of the double-well energy potential for a ferromagnet at $\mu_0 H_z = H_c$, representing two stable states corresponding to the magnetization pointing up and down. (b) Schematic of the energy potential under an external magnetic field $|H_c| < |\mu_0 H_z| < |H_s|$. The blue and red arrows represent islands with magnetization pointing down and up, respectively. (c) Illustration of the dipolar field $\vec{B}_{dip(\vec{r})}$ exerted by island $m_i$ on its neighboring island $m_j$. The magnetic moment $\vec{m}$ points upward, and the distance vector $\vec{r}$ lies in the plane. An additional external magnetic field $\mu_0 H_z$ is applied downward.

In order to investigate the magnetic interaction between the islands, we fabricated arrays of $9 \times 9$ islands. The inter-island distance, $s$, was varied by controlling the intervening amorphous or etched area. The Ga⁺ FIB was employed to fabricate arrays with four distinct inter-island distances ($s = 60, 80, 100$, and $200$ nm), as shown in the STEM images presented in Figs. 3 a-d. Three arrays ($s = 60, 80,$ and $100$ nm) were fabricated from a flake with a thickness of $d = 35$ nm. The distance between the centers of the amorphous track in the arrays is kept constant (300 nm), resulting in an island width of $w = 300 - s$ nm. The array with $s = 200$ nm was defined by etching rather than amorphization, resulting in trapezoidal islands with a width of 150 nm width and a thickness of 60 nm (Fig. 4d). For this array, the distance between the lines was slightly larger (350 nm). Previous work demonstrated that amorphous CGT is non-magnetic[28]. This work confirms our previous result and no difference was observed between etched and amorphized grids.

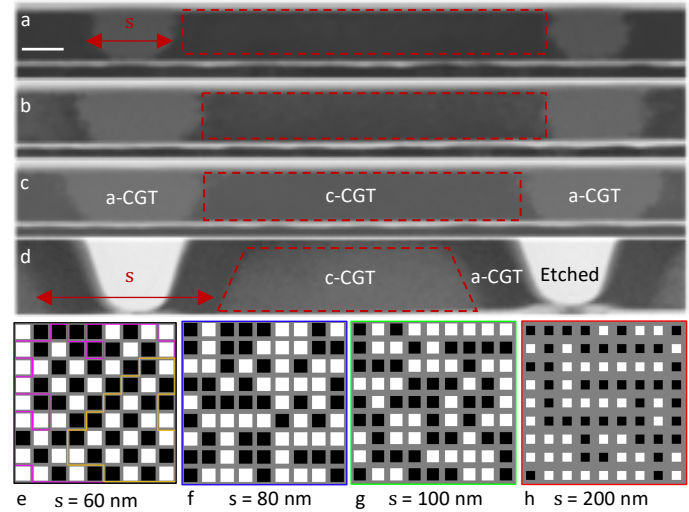

**Figure 4. Separation dependent inter-island interactions** (a-d) Cross-section of scanning transmission electron microscopy images of the CrGeTe₃ (CGT) arrays. The scale bar is 30 nm. (e-h) Sequence of binary matrices computed from the SQUID-on-tip (SOT) $B_z(x,y)$ images for island separations of $s = 60$ **e**, 80 **f**, 100 **g**, and 200 **h** nm. The purple and gold lines in **e** delineate two fully anticorrelated sublattices. The magnetization curves $M(H_z)$ of the nano-patterned arrays is presented in Supplementary Figure 1. The black/white color scale represents the magnetic moments pointing down/up, and the gray represents amorphous CGT. The SOT images were acquired at the relevant coercive fields, $\mu_0 H_c = 68$ **e**, 89 **f**, 99 **g**, and 70 **h** mT.

Figs. 4e-h depict the binary representation of the magnetic arrays at the coercive field, $H_c$, where $M(H_z = H_c) \sim 0$. The emerging magnetic patterns for $s = 60$ nm (Fig. 4e), reveal a propensity for each island to orient the magnetization in a direction antiparallel to its neighbors. Conversely, as observed in Fig. 4h ($s = 200$ nm), more distant neighboring islands are more often magnetized in parallel to each other, forming larger magnetic domains. The field

evolution of the arrays is presented in Supplementary Movies 3-6. In the discussion, we analyze the correlation and consider the relevant energy scales of the system.

**Discussion**

We now quantify the strength of this AFM correlation as a function of the inter-island separation, $s$. To determine the spatial autocorrelation within the matrices depicted in Fig. 4, we employ the Moran's $I$ metrics[30,31]. $I$ is calculated from the following expression:

$$I = \frac{n}{\sum_i \sum_j w_{i,j}} \frac{\sum_i \sum_j w_{i,j} (m_i - \bar{M})(m_j - \bar{M})}{\sum_i (m_i - \bar{M})^2}. \quad (1)$$

where, $n$ is the number of islands, $m_i, m_j$ refer to the island magnetization in the location $i, j$, $\bar{M} = M(H_z)/M_{Sat}$ is the normalized mean value of the array magnetization, and $w_{i,j}$ is the spatial weight between cells. To consider the nearest neighbor correlation, we set $w_{i,j} = 1$ between nearest neighbors, while all the other $n^2 \times n^2$ matrix entries are set to zero. The $I$ values range between -1 and 1, with a value of -1 indicating perfect negative correlation and a dominant AFM interaction (Fig. 5a, left panel). $I = 0$ indicates the absence of any correlation, as expected for a randomly distributed matrix (Fig. 5a, central panel). A value of 1 is obtained in the case of perfect positive correlation (ferromagnetic interaction), which would result in two magnetic domains pointing in opposite directions (Fig. 5a, right panel). We note that this metric is also useful away from the coercive field but is not defined for a fully magnetized array ($M(H_z) = M_{Sat}$) since the denominator, $\sum_i (m_i - \bar{M})^2$, vanishes when $m_i = \bar{M} \, \forall \, i$.

Fig. 5b presents a distribution of Moran's $I$ values derived from computer-generated $9 \times 9$ random matrices at $H_c$ ($M(H_z = H_c) = 0$) and based on a set of $10^4$ trials (gray distribution). This distribution is centered around $I = 0$, highlighting zero spatial correlation in the absence of dipolar interactions. Each marker plotted in Fig. 5b represents the experimentally determined mean $I$ values at $H_c$, obtained from distinct field sweeps. The error bars represent the obtained standard deviation. For the largest separation ($s = 200$ nm), we obtain $I = 0.00 \pm 0.08$, which indicates negligible interaction (Fig. 5b, red marker). Conversely, significant anticorrelations between the islands are observed for $s < 100$ nm ($I = -0.08 \pm 0.05, -0.16 \pm 0.08,$ and $-0.25 \pm 0.1$, for $s = 100, 80,$ and $60$ nm, respectively).

Fig. 5c presents the Moran's $I$ values as a function of the magnetization, $M(H_z)$ of the array. The gray points indicate values derived from random matrices, based on $10^3$ trials for each $M(H_z)$ value. We restrict ourselves to $|M(H_z)|/M_{Sat} < 0.8$ because the metric diverges at $|M(H_z)|/M_{Sat} = 1$. For $s = 200$ nm, the interaction appears weak ($I \approx 0$) for all $M(H_z)$ values, which resembles the situation in the random case. For $s = 60$ nm, at ($|M(H_z)|/M_{Sat} \approx 0.8$), we obtain $I \approx 0$, which suggests little anticorrelation for a relatively large magnetization state. In contrast, low magnetization, $|M(H_z)|/M_{Sat} < 0.8$, is associated with clearly negative $I$ values, indicating that dipolar interactions induce AFM correlations that vanish at stronger applied magnetic fields. We have also computed and plotted Moran's $I$ for an array of $23 \times 23$ islands with a separation of $s = 70$ nm, as shown in Fig. 2 (Fig. 5c, magenta). As anticipated, the array exhibits a substantial interaction, albeit weaker than that observed for a separation of $s = 60$ nm.

The anticorrelations revealed by our results are significant but not dominant since $I < 0.3$. This can be attributed to randomizing effects such as finite-temperature fluctuations or inhomogeneous island properties. These fluctuations could explain the finite range of fields at which the magnetization reverses. In absence of inhomogeneity and thermal fluctuations all islands should reverse their magnetization at the same field. In contrast, we observe consistently a transition over a field range of 70 mT (Fig. 2g and Supplementary Figure 1). Another source of disorder is the two-fold degeneracy of the AFM lattice (Fig. 4e). In previous works, the application of demagnetization[32,33] or thermal[34] protocols were used to relax the system into the ground state. This is visible in Fig. 4e where the top right part and the bottom left part are locked in different AFM domains. More than 80% of the nanoislands are part of these two domains and most of the remaining nanoisland are located at the grid's edge where there is a reduced number of neighbors.

The effect of thermal fluctuations is also noticeable when imaging the grid at the same field for an extended period of time. The effect of thermal activation is visible in Supplementary Movie 7, where we measure at a fixed field every 15 minutes during a six-hour period. We notice that during that time, the magnetization evolved from $M/M_{tot} = -0.4$ to $-0.1$ (Supplementary Figure 2). Thermal relaxation also allows the grid to approach a more anticorrelated state as revealed by the value of Moran's $I$, which evolves from $-0.1$ to $-0.17$. The thermal activation at 4.2 K suggests that thermal annealing procedure would be beneficial to enhance anticorrelations. It demonstrates that thermal activation is a relevant energy scale at 4.2 K near the coercive field. However, at zero applied field the anisotropy barrier is too large (10 eV) and no thermal activation was observed.

To determine the strength of interaction, we used the finite element method to calculate the dipolar energy (see methods), $E_{dip}$, between two islands with separation $s$ (Fig. 5d). The results reveal $E_{dip}$ grows by two orders of magnitude between $s = 200$ nm and $s = 60$ nm. This calculation explains why the Moran's $I$ grows significantly with decreasing $s$. However, it cannot explain the observed variation in $H_c$ since $K \gtrsim 10$ eV and $E_{dip} \leq 0.8$ eV (Supplementary Table 1). The measured variations in $H_c$ depend on the island dimensions and not on the spacing $s$ as discussed elsewhere.[35] Improving the uniformity of the island properties and enhancing $E_{dip}$ is achievable by using a higher resolution amorphization tool such as He FIB[36]. This would also allow us to decrease the separation to $s = 10$ nm, which would increase the dipolar energy by nearly an order of magnitude.

To explore which kind of response would be physically expected at smaller spatial scales than those achievable experimentally, we have generated squared arrays of CGT-based islands of dimension $23 \times 23 \times 3$ nm$^3$ in the framework of atomistic spin dynamics simulations [37–40] (see methods). The selected dimensionality of the modeled flattened islands maintains the same aspect ratio as in the experimental case but scaled down by an overall factor of ~10. We initially left the system to equilibrate by $5 \cdot 10^5$ Monte Carlo steps at a constant temperature of $T = 90$ K, well above the simulated Curie temperature ($T_C = 61.1$ K). Afterwards, a zero-field cooling process was carried out for $10^6$ steps, where the temperature was linearly reduced from an initial state ($T = 90$ K) to the targeted final value ($T = 0$ K). Once this stage was reached, the system was allowed to evolve for $2.5 \cdot 10^5$ extra steps with the aim of ensuring good convergence in the final state. Throughout the entire process, the demagnetizing contribution has been characterized through the inter-intra dipole tensor approach aiming for computational efficiency[41].

Within this computational framework, we preliminarily explore the emergence of anticorrelation in arrays with different numbers of CGT-based magnetic islands separated by a distance ($s$). Interestingly, we found that for grids of a $3 \times 3$ array, or smaller, no sizable dipolar-induced AFM interactions between the islands were present. The systems stabilized in a random orientation despite of the initial spin states and magnitudes of $s$. This is attributed to the fact that in a small array, a large proportion of the island are located at the edge of the array and thus have fewer neighbors. Similar effect was observed experimentally at the edge of larger array as discussed above (Fig. 5e). We observe that anticorrelation effects start to be present at systems composed by squared $5 \times 5$ arrays with island dimension of $23 \times 23 \times 3$ nm$^3$. In this scenario, an array with $s = 1, 5, 10$, and 20 nm (Fig. 5e) clearly displays an increasing AFM coupling between the islands as $s$ decreases. The Moran's $I$ index computed for each configuration (Fig. 5f) followed the same trend observed in the experiments (Fig. 5b) at a much smaller spatial scale. We note that the anticorrelation is significantly larger than in experiments due to the small spacing. This indicates that similar phenomena are present in both situations. Furthermore, the spatial distribution of the normalized out-of-plane $z$-th demagnetization field component ($B_z^{\text{demag}}$) for $s = 10$ nm (Fig. 5g) remarkably demonstrated that dipolar interactions are the main driving-force for the anticorrelation across the magnetic grid. Neighboring islands with a parallel magnetization (e.g., blue-blue or red-red) tend to be ferromagnetically coupled, whereas anti-parallel configurations (e.g., blue-red) shows that the demagnetizing field canceled out in the middle region between the islands. In the limit of a full anticorrelation (Moran's $I \sim -1$), the magnetic grid would develop zero stray fields ($B_z^{\text{demag}} = 0$) between the islands with resemblance of a chess-board pattern (Supplementary Figure 3). Simulations at higher temperature resulted in a $I$ closer to zero highlighting the importance of minimizing thermal fluctuations (Supplementary Figure 4).

In conclusion, we have demonstrated that by nanopatterning a 2D vdW magnet in a square grid we can finely adjust interactions to create states not accessible in isolated pristine flakes, which exhibit zero net magnetization at zero field. Such lab-made structures provide an exciting playground to investigate a wide class of fundamental correlated phenomena in CGT and in other 2D vdW magnets. We note that the dipolar coupling generates a rich type of interactions. For example, in a material with in-plane anisotropy would result in a ferromagnetic interaction along the magnetic moment and an antiferromagnetic interaction along the perpendicular direction. Further investigations could combine the effective antiferromagnetic interactions, mediated by the dipolar field with other lattices that induce geometric frustration to study artificial spin ice in magnetic 2D materials. These phenomena are not beyond reach thanks to lithography techniques, and the flexibility in sample fabrication using vdW material technology. As opposed to vertical approaches, commonly used through stacking layers on top of each other following different protocols (e.g., moiré stacking, multiple assembly, intercalation), the lateral engineering of magnetic layers provides an intriguing platform to generate disparate structures with tunable behavior. Our results open pathways to connect underlying results in spin-ice systems, frustration physics, and criticality to the ultimate thin limit routinely achieved in 2D materials but not so in more traditional magnets.

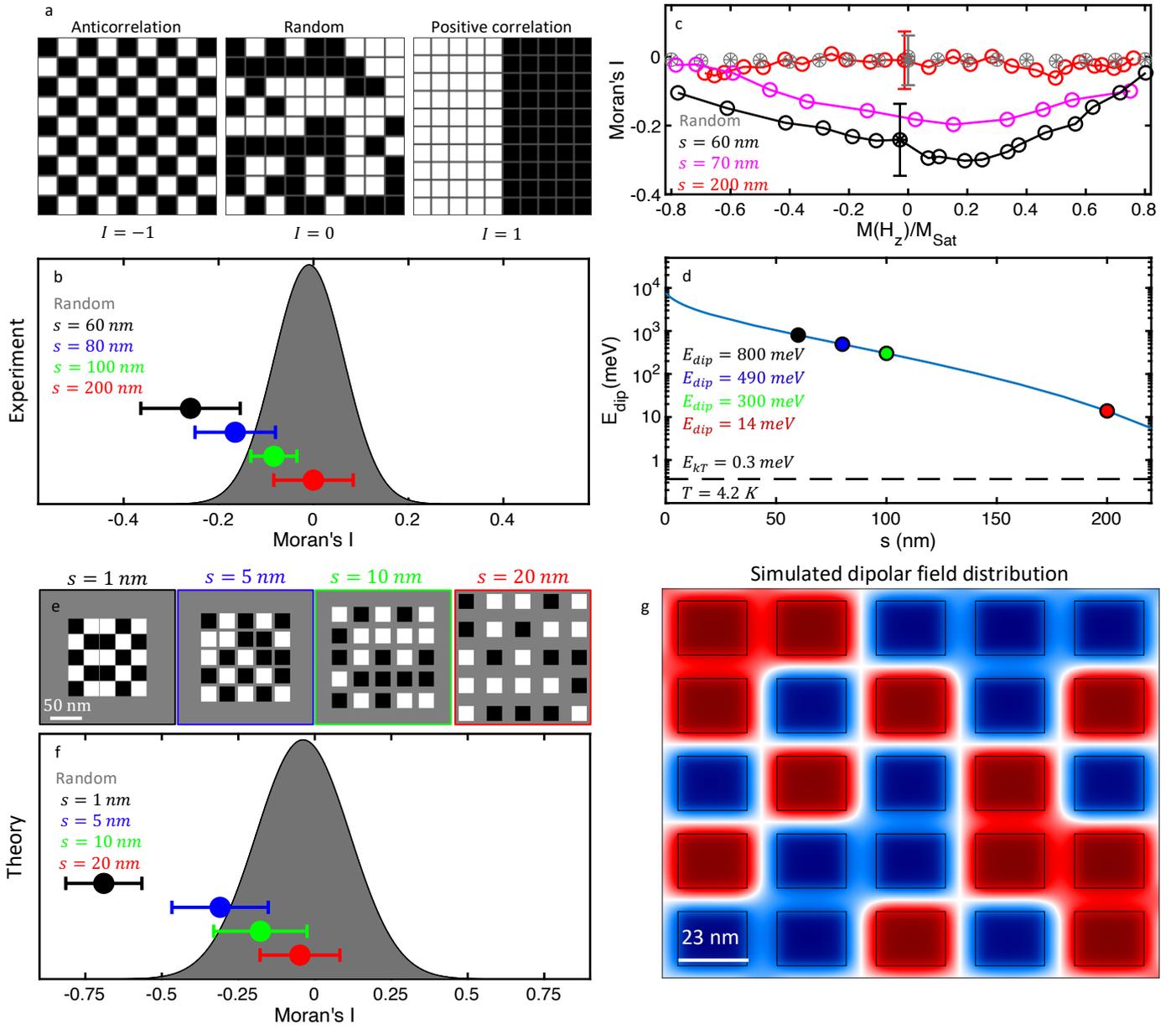

**Figure 5. Quantitative analysis of the correlation for arrays with distinct separations, $s$** (a) Illustration of matrices at the coercive field ($M(H_z = H_c) = 0$) with different types of inter-island correlations. A checkboard matrix that shows perfect anticorrelation with Moran's $I = -1$ (left). A random matrix, which has zero spatial correlation $I = 0$ (center). A perfectly correlated matrix for which $I = 1$. (b) Gray: distribution of Moran's $I$ values calculated for random $9 \times 9$ matrices at the coercive field based on a set of $10^4$ trials. Distinct markers represent the average Moran's $I$ measured for CrGeTe$_3$ (CGT) arrays at $H_c$ for separations of $s = $ 60 (black), 80 (blue), 100 (green), and 200 (red) nm. The markers indicate the mean values, and the error bars the range within one standard deviation. The number of repetitions varies for each sample, being 8, 8, 7, and 5, respectively. $I = 0.00 \pm 0.08, -0.08 \pm 0.05, -0.16 \pm 0.08,$ and $-0.25 \pm 0.1$ for $s = 100, 80$ and 60 nm, respectively. (c) Magnetization dependence of Moran's $I$ value for $s = 60$ (black), $s = 70$ (magenta), 200 nm (red), and computer-generated random matrices (grey). The magnetization is normalized to that of the fully magnetized array, $M_{Sat}$. The number of repetitions for each curve is 8 (red), 1 (magenta), and 3 (black) loops. The results for $s = 70$ nm are taken from Fig. 2. The error bars are representative of all data points measured for a given array. (d) The dipolar energy, $E_{dip}$, (light blue line) between two islands with a separation $s$ obtained from finite element calculations. A constant distance of 300 nm was assumed, resulting in an island size of $w = 300 \text{ nm} - s$ and a thickness of $d = 35$ nm. The thermal energy, $E_{kT}$, at $T = 4.2$ K is included for reference (dashed black line). (e-g) Atomistic spin dynamics simulations for 5×5 CGT arrays with distinct separations $s$. (e) Illustration of matrices at the coercive field ($M(H_z = H_c) = 0$) with island dimension of $23 \times 23 \times 3$ nm$^3$ and $s = 1, 5, 10,$ and 20 nm. The black/white color scale represents the magnetic moments pointing down/up and the gray represents nonmagnetic CGT. (f) Gray: distribution of Moran's $I$ values calculated for random $5 \times 5$ matrices at the coercive field based on a set of $10^4$ trials. Distinct markers represent the average Moran's $I$ values of CGT arrays at $H_c$ for separations of $s = 1$ (black), 5 (blue), 10 (green), and 20 (red) nm. The markers indicate the mean values, and the error bars the range within one standard deviation. Each data point represents the average of ten iterations. $I = -0.05 \pm 0.1, -0.2 \pm 0.2, -0.3 \pm 0.2,$ and $-0.7 \pm 0.1$) for $s = 20, 10, 5$ and 1 nm respectively (g) Simulated spatial distribution of the $z$-th component of the demagnetizing field, $B_z^{demag}$, for a $5 \times 5$ squared array composed of rectangular grains of $23 \times 23 \times 3$ nm$^3$ for an inter-island separation of $s = 10$ nm, which depicts a remarkable anticorrelation.

## Methods

Sample synthesis:

Single crystals of CrGeTe$_3$ were grown using slight excess of Te[42,43]. We grew our samples from high-purity Cr (Alfa Aesar 99.999%), Ge (GoodFellow 99.999%) and Te (GoodFellow 99.999%). Cr, Ge and Te were introduced and sealed in quartz ampoules. Then, the ampoules were heated from room temperature to 930 ºC in 12 h, cooled down to 715ºC in 54 h, and finally cooled down to 500 ºC in 54 h; here the samples remain during 99 h. We quenched the crystals down to ambient temperature by immersion in cold water. We obtained layered crystals of 2 mm × 2 mm × 0.5 mm.

Sample fabrication:

CGT samples were fabricated using the dry transfer technique, which was carried out in a glovebox with an argon atmosphere. The CGT flakes were cleaved using the scotch tape method and exfoliated on commercially available Gelfilm from Gelpack. The CGT flakes were transferred onto a SiO$_2$ substrate. The flakes were exfoliated from the crystals in areas without any Te flux. This was achieved by optically checking that the sample area was free of any inclusions and had large and flat surfaces. The various island shapes were etched or amorphized using a 30 keV Ga$^+$ focused ion beam (FIB)[28].

**Scanning SQUID-On-Tip microscopy:**

The SOT was fabricated using self-aligned three-step thermal deposition of Pb at cryogenic temperatures, as described previously[26]. The SQUID-on-tip resolution is determined primarily by the largest of the following parameters: the SQUID loop diameter and the tip-to-sample distance. In this work, the island size was about 200 nm, therefore we used a tip diameter ranging from 145 to 175 nm and the distance from the sample was ranging from 50 to 200 nm. All measurements were performed at 4.2 K in a helium atmosphere at 1 to 10 mbar. The sample was exposed to air about one hour each time it was loaded in the microscope or to change tip. Samples are typically exposed a few hours without any noticeable change in their magnetic properties.

**Sample characterization:**

High-resolution scanning electron microscope (SEM) cross-section lamellas were prepared and imaged by Helios Nanolab 460F1 Lite focused ion beam (FIB) - Thermo Fisher Scientific. The site-specific thin lamella was extracted from the CGT patterns using FIB lift-out techniques[44]. STEM and Energy-Dispersive X-ray Spectroscopy (EDS) analyses were conducted using an Aberration Prob-Corrected S/TEM Themis Z G3 (Thermo Fisher Scientific) operated at 300 KV and equipped with a high-angle annular dark field detector (Fischione Instruments) and a Super-X EDS detection system (Thermo Fisher Scientific). See Supplementary Note 1 and Supplementary Figure 5 for more details.

**Atomistic spin dynamics methods:**

The modeling of the initial rhombohedral CGT pristine structure has been reduced to the characterization of the spatial hexagonal arrangement of the Cr-based atoms in the magnetic unit cell, whose three-dimensional dimensions are given by $a = 6.91$ Å, $b = 11.97$ Å, and $c = 21.82$ Å. In order to successfully capture the physical response of the system, the configurational spin Hamiltonian, $\mathcal{H}$, includes:

$$\mathcal{H} = -\frac{1}{2}\sum_{i \neq j} \mathbf{m}_i\, \mathcal{J}_{ij}\mathbf{m}_j - K\sum_i (\mathbf{m}_i \cdot \hat{\mathbf{z}})^2 - \mu_s \sum_i \left(\mathbf{m}_i \cdot \mathbf{B}_i^{\text{demag}}\right), \quad (2)$$

where $\mathbf{m}_i$ and $\mathbf{m}_j$ represent the magnetization vectors at the $i$- and $j$-th atomic positions, normalized by the atomic magnetic moment, $\mu_s$. The symmetric exchange parameter, $\mathcal{J}_{ij}$, accounts for the bilinear intra- ($\mathcal{J}_1 = 7.52$ meV, $\mathcal{J}_2 = -0.16$ meV, and $\mathcal{J}_3 = 0.32$ meV), and interlayer ($\mathcal{J}_{z1} = -0.10$ meV, $\mathcal{J}_{z2} = 0.24$ meV, and $\mathcal{J}_{z3} = 0.76$ meV) interactions up to the third nearest neighbors were extracted from previous simulations[45]. The preferential direction for the magnetization in this material is along the $z$-th axis with a uniaxial single-ion anisotropy $K = 0.05$ meV[25]. The third term on Eq. (2) defines the magnetic dipolar contribution, an expression which is dominated by the demagnetizing field acting on the $i$-th magnetic atom, $\mathbf{B}_i^{\text{demag}}$. In this approach the calculation of the dipole interaction acting in the $i$-th magnetic atom is computed atomistically within a selected cut-off radius ($r_c$) centered on the reference atom, while to find the influence of atoms outside this range the macrocell formalism is followed. The micromagnetic cuboid cells for the long-range scenario have a fixed size, calculating for each of them their associated total magnetic moment through which they will act as effective dipoles for distant magnetic atoms. Thus, the demagnetizing field, $\mathbf{B}_i^{\text{demag}}$, included in Eq. (2), will be calculated based on the distance of each magnetic atom to the $i$-th reference one, $|\mathbf{r}_{ij}|$, as follows:

$$\mathbf{B}_i^{\text{demag}} = \begin{cases} \dfrac{\mu_0}{4\pi} \sum_{i \neq j} \dfrac{3(\mathbf{m}_i \cdot \hat{\mathbf{r}}_{ij})(\mathbf{m}_j \cdot \hat{\mathbf{r}}_{ij}) - (\mathbf{m}_i \cdot \mathbf{m}_j)}{|\mathbf{r}_{ij}|^3}, & \text{if } |\mathbf{r}_{ij}| \leq r_c, \\ \dfrac{\mu_0}{4\pi} \sum_{i \neq p} \dfrac{3(\mathbf{m}_i \cdot \hat{\mathbf{r}}_{ip})(\mathbf{m}_p^{\text{mc}} \cdot \hat{\mathbf{r}}_{ip}) - (\mathbf{m}_i \cdot \mathbf{m}_p^{\text{mc}})}{|\mathbf{r}_{ip}|^3}, & \text{if } |\mathbf{r}_{ij}| > r_c, \end{cases} \quad (3)$$

where $\mu_0$ represents the vacuum magnetic permeability, $\mathbf{m}_p^{\text{mc}}$ the magnetic moment of the $p$-th macrocell, and $\hat{\mathbf{r}}_{ip}$ the unit vector linking the $i$-th magnetic atom and the center of mass of the $p$-th macrocell[37,45]. In our simulations, we have operated with a macrocell size ($1 \times 1 \times 1$ nm$^3$) and cut-off radius, which governs the transition from the atomistic to the macrocell dipolar characterization for atoms beyond 2 macrocells from the reference $i$-th magnetic atom.

**Finite element calculation**

The energy resulting from the dipolar interaction was calculated by the finite element method. The sample was assumed to be perfectly confined in the $xy$ plane. Two islands of width $w$ separated by $s$ were considered to calculate the nearest neighbors interaction. We divided each island into $N^2$ elements of $\frac{w}{N} \times \frac{w}{N}$. For each element $i$ of one island, we calculate the field generated on that element by all the other elements of the island ($\vec{B}_i = \frac{1}{4\pi} \sum_{i=1}^{N^2} \frac{-\vec{m}_j}{|r_{ij}|^3}$). The interaction energy is obtained as follow $E_i = -\vec{m}_i \cdot \vec{B}_i$. The total energy is obtained by summing the energy calculated for each element $E_{island} = \sum_{i=1}^{N^2} E_i$.

**Acknowledgements:**

We would like to thank A. Capua, E. Katzav and N. Katz for fruitful discussions. We thank A. Vakahi and S. Remennik for technical support. R.R.-E. and H.S. acknowledge the helpful comments made by S. Jenkins for the computational implementation of the system. This work was supported by the European Research Council (ERC) Foundation grant No. 802952 and the Israel Science Foundation (ISF) Grant No. 645/23. The international collaboration on this work was fostered by the EU-COST Action CA21144. H. Steinberg acknowledges funding provided by the DFG Priority program grant 443404566 and Israel Science Foundation (ISF) grant 861/19. O. Millo is grateful for support from the Academia Sinica – Hebrew University Research Program, the ISF grant no. 576/21, and the Harry de Jur Chair in Applied Science. H. Suderow and E. Herrera acknowledge support from the Spanish State Research Agency (PID2020-114071RB-I00, CEX2018-000805-M) and the Comunidad de Madrid through the NANOMAGCOST-CM program (Program No.S2018/NMT-4321). E.J.G.S. acknowledges computational resources through CIRRUS Tier-2 HPC Service (ec131 Cirrus Project) at EPCC (http://www.cirrus.ac.uk), which is funded by the University of Edinburgh and EPSRC (EP/P020267/1); and ARCHER2 UK National Supercomputing Service via the UKCP consortium (Project e89) funded by EPSRC grant ref EP/X035891/1. E.J.G.S. acknowledges the EPSRC Open Fellowship (EP/T021578/1) and the Donostia International Physics Center for funding support.

**Author Contributions:**

Y.A., and A.N. conceived the experiment.
E.H. and H.S. synthesized the CGT crystals.
A.N., Y.Z., and N.F. carried out the scanning SOT measurements.
Y.A., M.K., H. Steinberg, and A.N. fabricated the CGT devices.
A.N. characterized the CGT devices.
A.N., M.M., Y.K.K., T.L., and M.R.A. analyzed the data.
Y.A., and A.N. constructed the scanning SOT microscope.
M.E.H. developed the SOT readout system.
R.R.-E., H. Solanki, E.J.G.S., and S.G. provided theoretical and simulation support.
A.N., E.J.G.S., O.M., and Y.A. wrote the paper with contributions from all authors.
Notes: The authors declare no competing financial interest.

# Supplementary Material

*Field-induced antiferromagnetic correlations in a nanopatterned van der Waals ferromagnet: a potential artificial spin ice*

*Avia Noah\*[1,2,3], Nofar Fridman[1,2], Yishay Zur[1,2], Maya Markman[1], Yotam Katz King[1,2], Maya Klang[1], Ricardo Rama-Eiroa[4,5], Harshvardhan Solanki[5], Michael L. Reichenberg Ashby[1,6], Tamar Levin[1], Edwin Herrera[7], Martin E. Huber[8], Snir Gazit[1], Elton J. G. Santos\*[4,5,9], Hermann Suderow[7], Hadar Steinberg[1,2], Oded Millo[1,2], and Yonathan Anahory\*[1,2]*

[1]The Racah Institute of Physics, The Hebrew University, Jerusalem, 9190401, Israel
[2]Center for Nanoscience and Nanotechnology, Hebrew University of Jerusalem, Jerusalem, 91904, Israel
[3]Faculty of Engineering, Ruppin Academic Center, Emek-Hefer, 40250 Monash, Israel
[4]Donostia International Physics Center (DIPC), 20018 Donostia-San Sebastián, Basque Country, Spain
[5]Institute for Condensed Matter Physics and Complex Systems, School of Physics and Astronomy, University of Edinburgh, Edinburgh, EH93FD, United Kingdom
[6]Imperial College London, Blackett Laboratory, London, SW7 2AZ, United Kingdom
[7]Laboratorio de Bajas Temperaturas, Unidad Asociada UAM/CSIC, Departamento de Física de la Materia Condensada, Instituto Nicolás Cabrera and Condensed Matter Physics Center (IFIMAC), Universidad Autónoma de Madrid, E-28049 Madrid, Spain
[8]Departments of Physics and Electrical Engineering, University of Colorado Denver, Denver, CO 80217, USA
[9]Higgs Centre for Theoretical Physics, University of Edinburgh, Edinburgh EH93FD, United Kingdom

Email: avia.noah@mail.huji.ac.il, esantos@ed.ac.uk, yonathan.anahory@mail.huji.ac.il

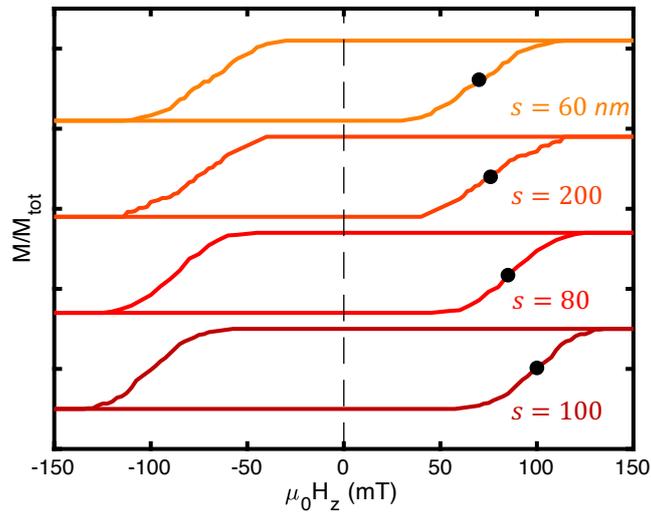

**Supplementary Figure 1 - Field evolution of island arrays in CrGeTe₃.** Hysteresis curves drawn from $B_z(x,y)$ measured on arrays with separation ranging between 60 and 200 nm. The array's coercive field $H_c^a$ is marked with black dots. The hysteresis curves were measured by ramping the field in one direction and were symmetrized to obtain the second branch of the $M(H_z)/M_{tot}$ curve. Curves were shifted vertically for clarity.

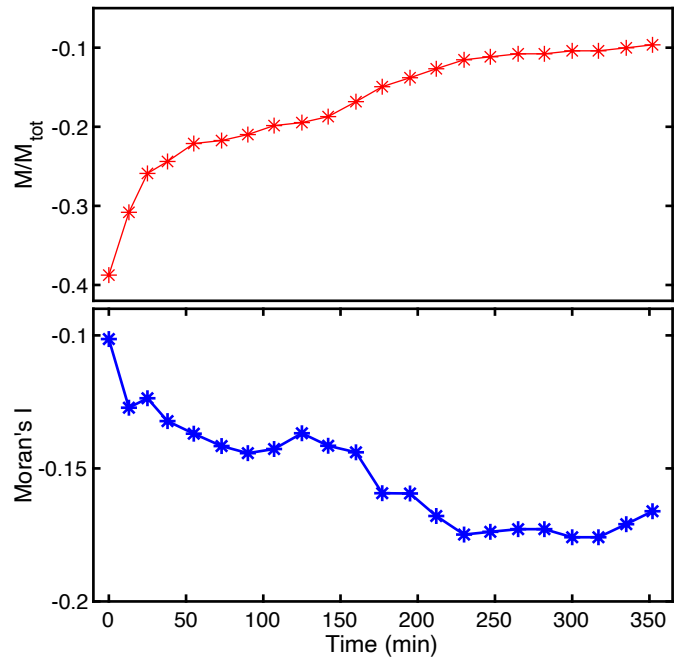

**Supplementary Figure 2 - Thermal activation of CrGeTe₃ array.** (a) Time evolution of the array magnetization at $\mu_0 H_z = 95$ mT. (b) Moran's I over time at $\mu_0 H_z = 95$ mT. The data corresponds to the array shown in Figure 2, consisting of 30x30 islands with a separation of 70 nm. Measurements were taken by continuously scanning the same area, with an imaging speed of 12 minutes per image and a 5-minute pause between images after the first four images (total of 22 images).

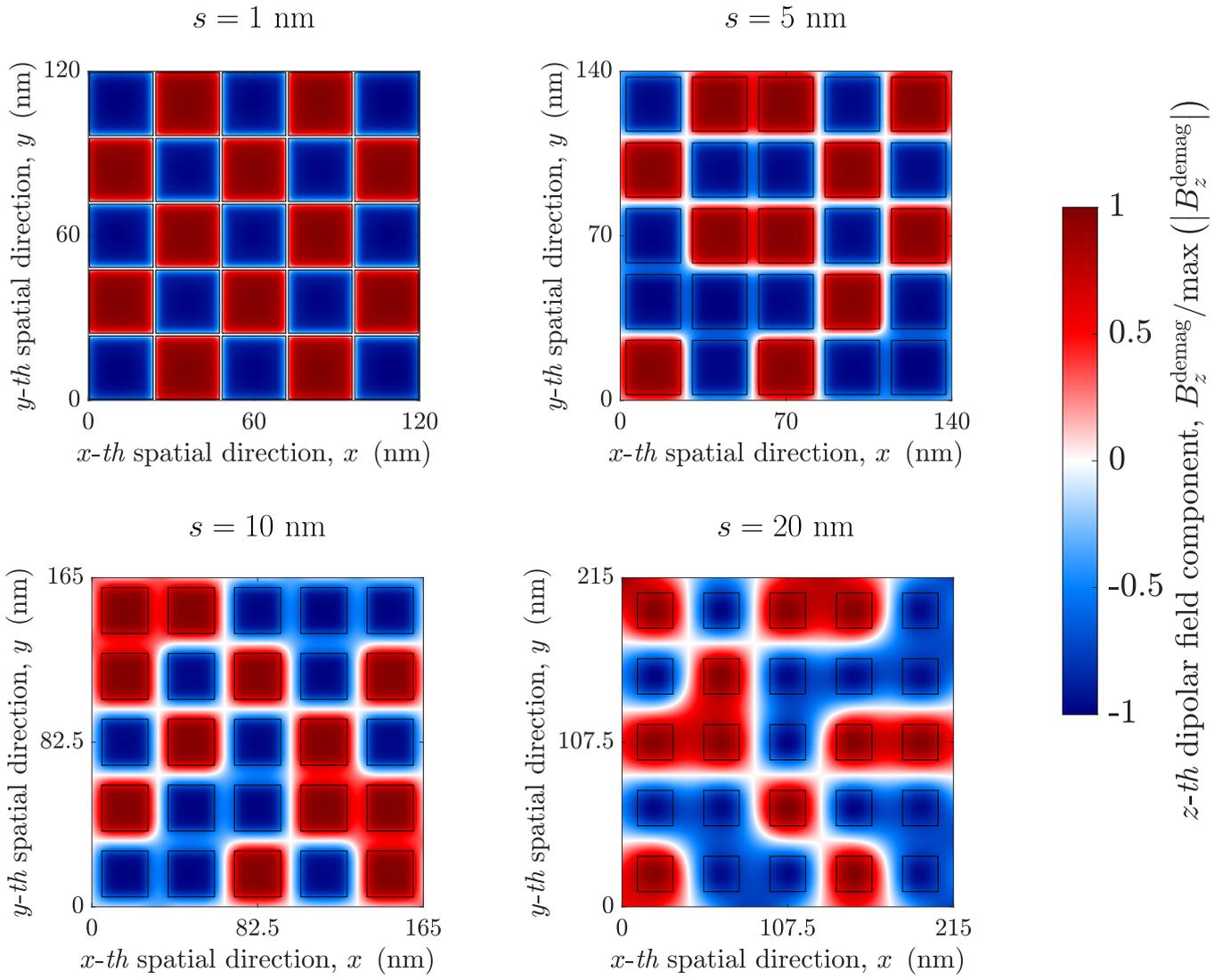

**Supplementary Figure 3 -** Simulated spatial distribution of the out-of-plane $z$-th component of the demagnetizing field, $B_z^{\text{demag}}$, for a $5 \times 5$ squared array of $23 \times 23 \times 3$ nm$^3$ rectangular islands for different inter-grain separations $s = 1, 5, 10,$ and $20$ nm after a zero-field cooling process at a final temperature of $T = 0$ K. The selected seeds included in the plots correspond to those closest to the mean Moran's $I$ value for each inter-island distance $s$ included in Figure 4f. We have employed a bicubic interpolation between the islands to obtain $B_z^{\text{demag}}$.

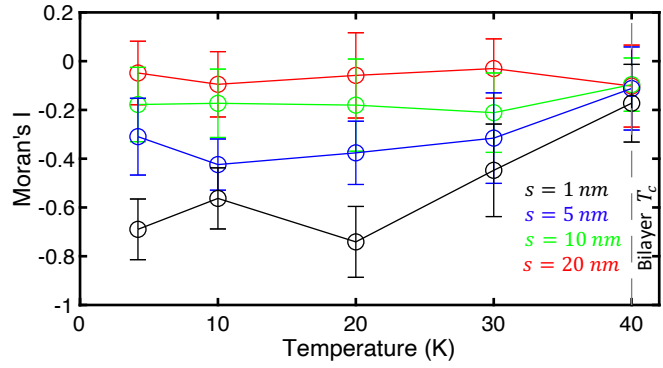

**Supplementary Figure 4 - Atomistic spin dynamics simulations of Moran's I Temperature dependent.** Distinct markers represent the average Moran's $I$ values of $5 \times 5$ CGT arrays at $H_c$ for separations of $s = 1$ (black), 5 (blue), 10 (green), and 20 (red) nm. The markers indicate the mean values, and the error bars the range within one standard deviation. The island dimension is $23 \times 23 \times 3$ nm$^3$. Each data point represents the average of 10 iterations.

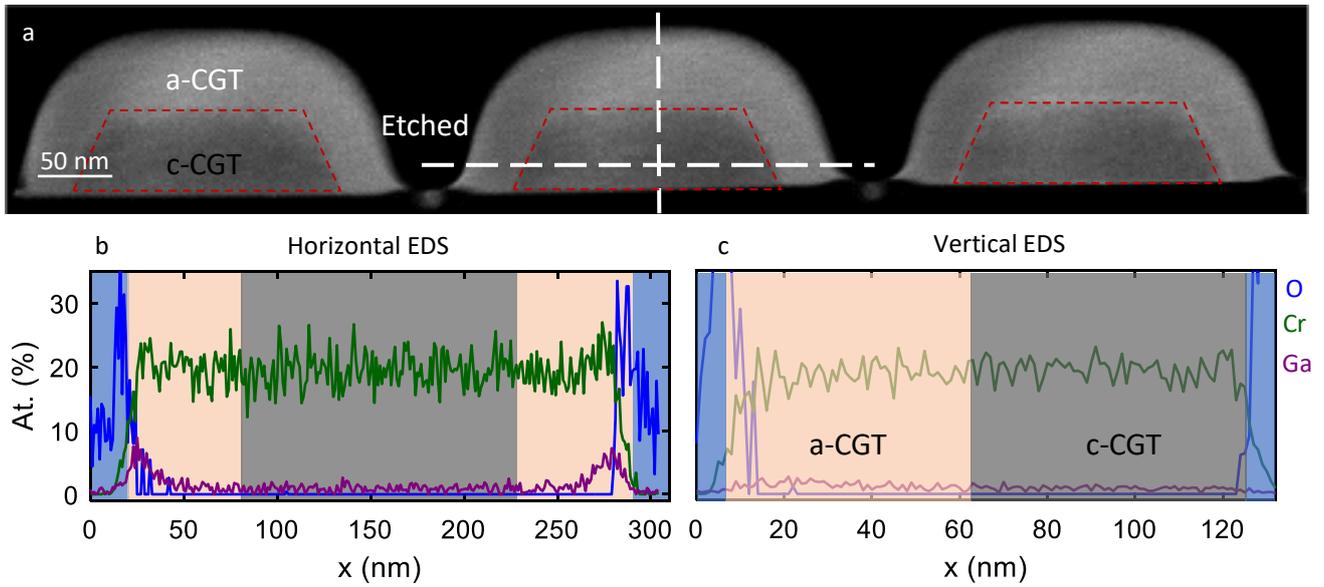

**Supplementary Figure 5 - Scanning transmission electron microscopy and Energy-Dispersive X-ray Spectroscopy analyses of CrGeTe$_3$ island.** (**a**) High-angle annular dark field (HAADF) image of the array with a separation of 200 nm. The area inside the dark red dotted line trapezoid is the region where magnetic crystalline CrGeTe3 is found. Above the crystalline region, an amorphous and non-magnetic region is observed. (**b-c**) Energy-Dispersive X-ray Spectroscopy (EDS), showing the relative amount of Cr, O, and Ga in the island cross-sections.

| Array # | Island Separation $s$ (nm) | Effective size $w \times w \times d$ (nm³) | Volume $V$ ($10^6$ nm³) | Island Magnetization $m_i = \dfrac{3\mu_b V}{V_{cell}}$ (eV T$^{-1}$) | Island Magnetization $m_i = \dfrac{3\mu_b V}{V_{cell}}$ ($10^6$ $\mu_b$) | Median island Coercivity $H_c$ (mT) | Dipolar Energy $E_{dip}$ (eV) | Median island anisotropy $K = \dfrac{H_c^i M}{2}$ (eV) |
|---|---|---|---|---|---|---|---|---|
| 1 | 60 ± 4 | 240 × 240 × 35 | 2.0 ± 0.1 | 420 ± 30 | 7.3 ± 0.5 | 67 ± 4 | 0.8 ± 0.1 | 14 ± 2 |
| 2 | 70 ± 4 | 230 × 230 × 30 | 1.6 ± 0.1 | 330 ± 20 | 5.7 ± 0.5 | 95 ± 4 | 0.5 ± 0.1 | 15 ± 2 |
| 3 | 80 ± 4 | 220 × 220 × 35 | 1.7 ± 0.1 | 360 ± 30 | 6.2 ± 0.5 | 89 ± 4 | 0.5 ± 0.1 | 15 ± 2 |
| 4 | 100 ± 4 | 200 × 200 × 35 | 1.4 ± 0.1 | 290 ± 20 | 5.0 ± 0.3 | 99 ± 4 | 0.3 ± 0.1 | 14 ± 2 |
| 5 | 200 ± 10 | 150 × 150 × 60 | 1.4 ± 0.1 | 280 ± 20 | 4.8 ± 0.3 | 70 ± 5 | 0.04 ± 0.1 | 10 ± 2 |

**Supplementary Table 1. A summary of the islands' parameters and results.** The uncertainty on the dimension is ±5 nm for the width $w$ and ±2 nm for the thickness $d$.

**Supplementary Note 1: Sample characterization:**

Scanning transmission electron microscopy images of CrGeTe$_3$ islands:

Lamellas were prepared and imaged by Helios Nanolab 460F1 Lite focused ion beam (FIB) - Thermo Fisher Scientific. The site-specific thin lamella was extracted from the CGT array using FIB lift-out techniques[1]. Scanning transmission electron microscopy (STEM) and Energy-Dispersive X-ray Spectroscopy (EDS) analyses were conducted using an Aberration Prob-Corrected S/TEM Themis Z G3 (Thermo Fisher Scientific) operated at 300 KV and equipped with a high-angle annular dark field (HAADF) detector from Fischione Instruments and a Super-X EDS detection system (Thermo Fisher Scientific). To determine the CGT islands' dimensions, we performed cross-section STEM on all islands of the main text. HAADF STEM images are shown in Figure 1e and Figure 3a-d.

Energy-Dispersive X-ray Spectroscopy of CrGeTe$_3$ island

In Supplementary Figure 5 we present the High-angle annular dark field (HAADF) image of the $150 \times 150 \times 60$ nm$^3$ island. The image resolves that the crystal structure is damaged due to the FIB etching. Near the etched area, the material is amorphous (bright gray color scale) where the crystalized CGT appears darker. The images reveal the precise thickness of the flake ($d = 60 \pm 2$ nm) and the edge cross-section $w = 150 \pm 5$ nm. To understand the stoichiometry of the flakes we perform an energy-dispersive spectroscopy (EDS) measurement. The EDS reveals accumulation of Ga and oxidation peaks near the amorphous edge. The concentration decays abruptly over a length of tens nm. We emphasize that Ga concentration peaks appear only in the amorphous part which we found to be non-magnetic. The Ga in the crystalline area less than 2% according to our EDS measurements. Traces of Silicon were also observed in the EDS measurements which seem to originate from organic residues from the PDMS used during the exfoliation process.

**References:**

[1]   M. Sezen, in *Modern Electron Microscopy in Physical and Life Sciences*, InTech, **2016**.

**Supplementary Movies:**

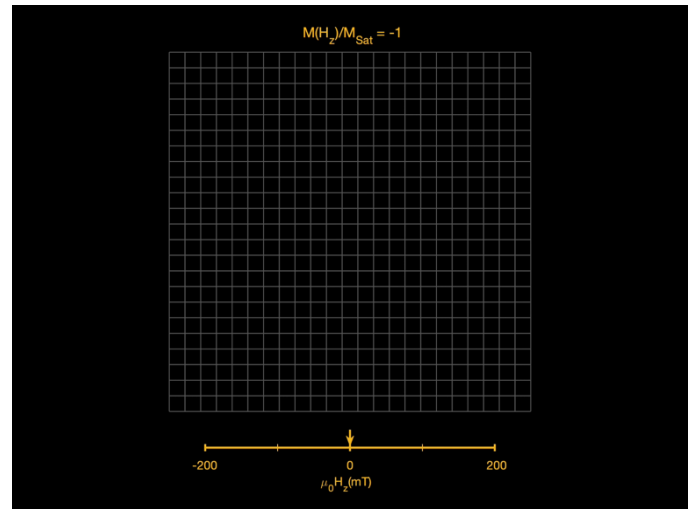

**Supplementary Movie 1 – Magnetic field response of the CrGeTe₃ (CGT) array magnetization.** The movie shows one magnetization loop of CGT nanoislands array, captured using SQUID-on-tip (SOT) $B_z(x,y)$ images at distinct values of the applied out-of-plane magnetic field $\mu_0 H_z$. The images were acquired between $\mu_0 H_z = \pm 200$ mT. The SOT image sizes are $8 \times 8$ μm². The black/white color scale represents the magnetic moments pointing antiparallel/parallel (down/up) to the applied field.

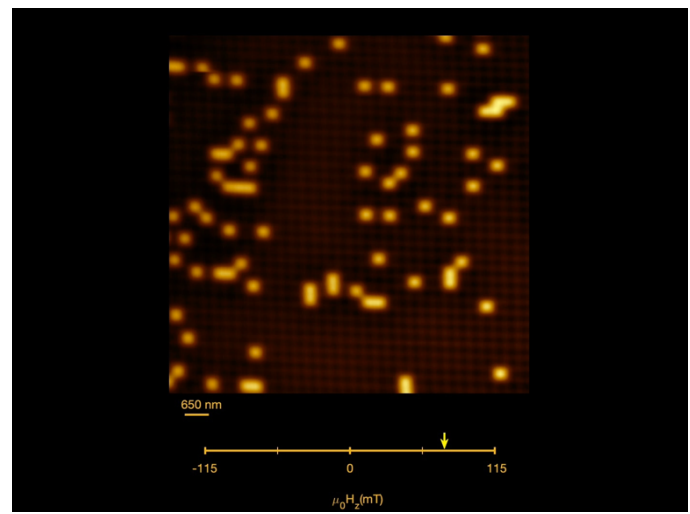

**Supplementary Movie 2 – Magnetic field response of the CrGeTe₃ (CGT) array magnetization.** The movie shows three magnetization loops (positive and negative) of CGT nanoislands array, captured using SQUID-on-tip (SOT) $B_z(x,y)$ images at distinct values of the applied out-of-plane magnetic field $\mu_0 H_z$. The images were acquired between $\mu_0 H_z = \pm 75$ mT and $\mu_0 H_z = \pm 115$ mT following a field excursion of $\mu_0 H_z = \mp 200$ mT. The SOT image sizes are $10 \times 10$ μm². The black/yellow color scale represents the magnetic moments pointing antiparallel/parallel (down/up) to the applied field.

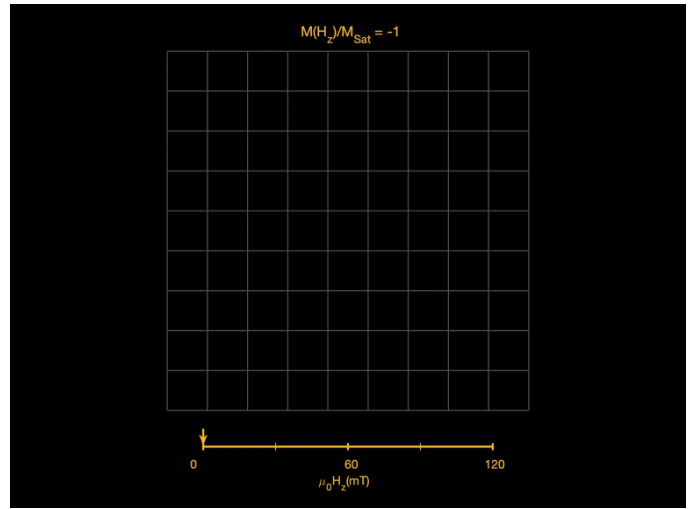

**Supplementary Movie 3 – Magnetic field response of the CrGeTe₃ (CGT) array with separation of 60 nm.** The movie shows the magnetic evolution of CGT nanoislands array, captured using SQUID-on-tip (SOT) $B_z(x,y)$ images at distinct values of the applied out-of-plane magnetic field $\mu_0 H_z$. The images were acquired between $\mu_0 H_z = 0$ to $\mu_0 H_z = 120$ mT. The SOT image sizes are $5 \times 5$ μm². The black/white color scale represents the magnetic moments pointing antiparallel/parallel (down/up) to the applied field.

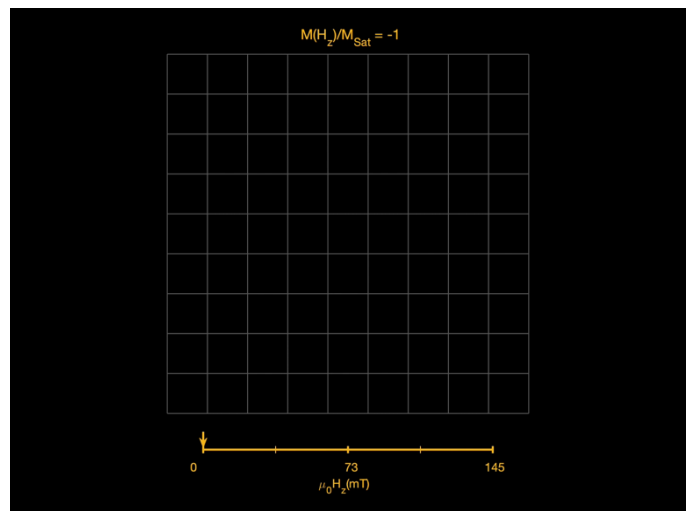

**Supplementary Movie 4 – Magnetic field response of the CrGeTe₃ (CGT) array with separation of 80 nm.** The movie shows the magnetic evolution of CGT nanoislands array, captured using SQUID-on-tip (SOT) $B_z(x,y)$ images at distinct values of the applied out-of-plane magnetic field $\mu_0 H_z$. The images were acquired between $\mu_0 H_z = 0$ to $\mu_0 H_z = 145$ mT. The SOT image sizes are $5 \times 5$μm². The black/white color scale represents the magnetic moments pointing antiparallel/parallel (down/up) to the applied field.

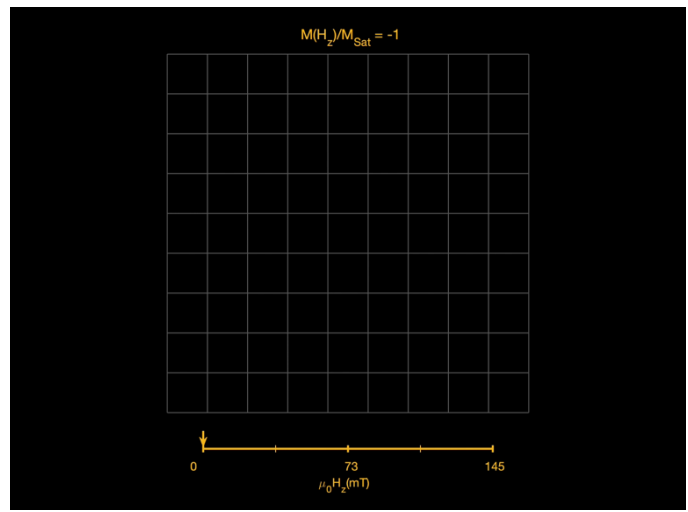

**Supplementary Movie 5 – Magnetic field response of the CrGeTe₃ (CGT) array with separation of 100 nm.** The movie shows the magnetic evolution of CGT nanoislands array, captured using SQUID-on-tip (SOT) $B_z(x,y)$ images at distinct values of the applied out-of-plane magnetic field $\mu_0 H_z$. The images were acquired between $\mu_0 H_z = 0$ to $\mu_0 H_z = 145$mT. The SOT image sizes are $5 \times 5$ μm². The black/white color scale represents the magnetic moments pointing antiparallel/parallel (down/up) to the applied field.

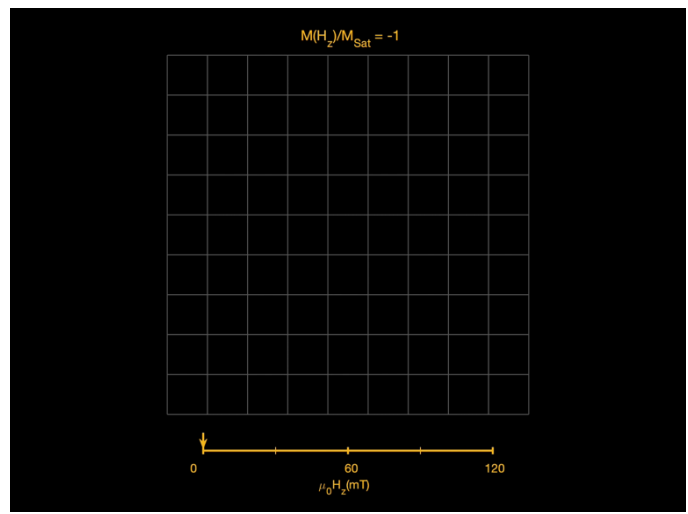

**Supplementary Movie 6 – Magnetic field response of the CrGeTe₃ (CGT) array with separation of 200 nm.** The movie shows the magnetic evolution of CGT nanoislands array, captured using SQUID-on-tip (SOT) $B_z(x,y)$ images at distinct values of the applied out-of-plane magnetic field $\mu_0 H_z$. The images were acquired between $\mu_0 H_z = 0$ to $\mu_0 H_z = 120$ mT. The SOT image sizes are $4 \times 4$ μm². The black/white color scale represents the magnetic moments pointing antiparallel/parallel (down/up) to the applied field.

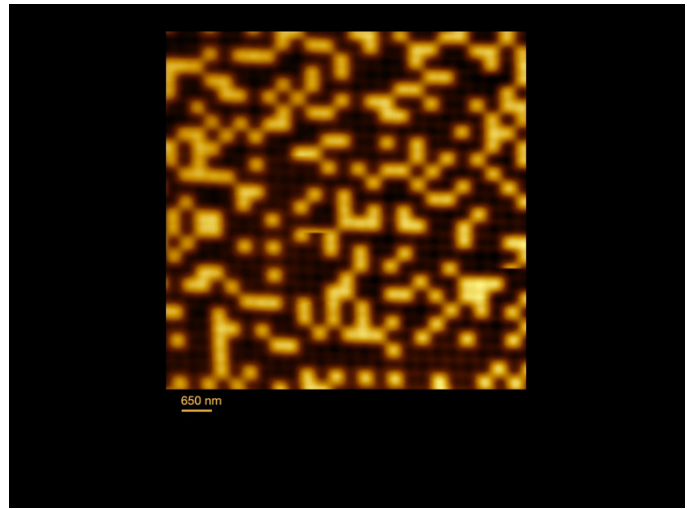

**Supplementary Movie 7 – Thermal activation of CrGeTe$_3$ array.** Time evolution of the array magnetization at $\mu_0 H_z = 95$ mT. The data corresponds to the array shown in Figure 2, consisting of 30x30 islands with a separation of 70 nm. Measurements were taken by continuously scanning the same area, with an imaging speed of 12 minutes per image and a 5-minute pause between images after the first four images (total of 22 images).